\documentclass[12pt]{article}
\usepackage{amsmath,epsf,amssymb,latexsym,cite,xy}
\xyoption{matrix}\xyoption{arrow}
\newtheorem{theorem}{Theorem}

\newif\iffigs\figstrue

\DeclareFontFamily{U}{rsf}{}
\DeclareFontShape{U}{rsf}{m}{n}{
  <5> <6> rsfs5 <7> <8> <9> rsfs7 <10-> rsfs10}{}
\DeclareMathAlphabet\Scr{U}{rsf}{m}{n}

\def\pplogo{\vbox{\kern-\headheight\kern -43pt
\halign{##&##\hfil\cr&{
\ppnumber}\cr\rule{0pt}{2.5ex}&\ppdate\cr}
}}
\makeatletter
\def\ps@firstpage{\ps@empty \def\@oddhead{\hss\pplogo}%
  \let\@evenhead\@oddhead 
}
\def\maketitle{\par
 \begingroup
 \def\thefootnote{\fnsymbol{footnote}}
 \def\@makefnmark{\hbox{$^{\@thefnmark}$\hss}}
 \if@twocolumn
 \twocolumn[\@maketitle]
 \else \newpage
 \global\@topnum\z@ \@maketitle \fi\thispagestyle{firstpage}\@thanks
 \endgroup
 \setcounter{footnote}{0}
 \let\maketitle\relax
 \let\@maketitle\relax
 \gdef\@thanks{}\gdef\@author{}\gdef\@title{}\let\thanks\relax}
\makeatother

\def\O{\Scr{O}}
\def\C{{\mathbb C}}
\def\P{{\mathbb P}}

\def\R{{\mathbb R}}
\def\Z{{\mathbb Z}}

\def\Hom{\operatorname{Hom}}
\def\hom{\operatorname{hom}}
\def\sHom{\operatorname{\Scr{H}\!\!\textit{om}}}
\def\Ext{\operatorname{Ext}}
\def\End{\operatorname{End}}

\def\tr{\operatorname{tr}}
\def\Tr{\operatorname{Tr}}

\def\Img{\operatorname{Im}}
\def\Rea{\operatorname{Re}}

\def\GU{\operatorname{U{}}}

\def\Cone{\operatorname{Cone}}

\def\ch{\operatorname{\mathit{ch}}}
\def\td{\operatorname{\mathit{td}}}

\def\sm{$\sigma$-model}
\def\nlsm{non-linear \sm}

\def\CY{Calabi--Yau}
\def\LG{Landau--Ginzburg}

\def\cI{{\Scr I}}

\def\cE{{\Scr E}}
\def\cF{{\Scr F}}

\def\cG{{\Scr G}}
\def\DC{\mathbf{D}}
\def\CT{\mathbf{T}}

\def\ie{{\em i.e.}}

\begin{document}
\setcounter{page}0
\def\ppnumber{\vbox{\baselineskip14pt
\hbox{DUKE-CGTP-01-07}
\hbox{SLAC-PUB-8812}
\hbox{SU-ITP-01/18}
\hbox{hep-th/0104147}}}
\def\ppdate{April 2001} \date{}

\title{\LARGE Derived Categories and Zero-Brane Stability\\[10mm]}
\author{
Paul S. Aspinwall\\[2mm]
\normalsize Center for Geometry and Theoretical Physics, \\
\normalsize Box 90318, \\
\normalsize Duke University, \\
\normalsize Durham, NC 27708-0318\\[5mm]
Albion Lawrence\\[2mm]
\normalsize Theoretical Physics Group, MS 81,\\
\normalsize Stanford Linear Accelerator Center,\\
\normalsize 2575 Sand Hill Road,\\
\normalsize Menlo Park, CA 94025;\\
\normalsize and\\
\normalsize Stanford University Physics Department,\\
\normalsize Stanford, CA 94305\\
}

{\hfuzz=10cm\maketitle}

\def\Large{\large}
\def\LARGE{\large\bf}


\begin{abstract}
We define a particular class of topological field theories associated
to open strings and prove the resulting D-branes and open strings form
the bounded derived category of coherent sheaves. This derivation is a variant
of some ideas proposed recently by Douglas. We then argue that any
0-brane on any \CY\ threefold must become unstable along some path in the
K\"ahler moduli space. As a byproduct of this analysis we see how the
derived category can be invariant under a birational transformation
between \CY s.
\end{abstract}

\vfil\break


\section{Introduction}    \label{s:intro}

The idea that a D-brane is simply some subspace of the target space
where open strings are allowed to end is clearly too simple. Even
at zero string coupling, we are faced with
carefully analyzing the \nlsm\ of maps from the string worldsheet into
the target space. It is well-known that the \nlsm\ modifies the usual
rules of classical geometry in the target space.
We expect that the notion of a D-brane should get quite complicated
once such notions of stringy geometry are taken into account.

One computational handle on this issue is via 
string-scale target spaces that have an exact CFT description.
There, D-branes can be explicitly constructed as ``boundary states''
after the fashion of \cite{CLNY:,JoeCai:}.
Such states have been studied at the Gepner point of a variety of
Calabi-Yau threefold compactifications.  Existing techniques
have yielded only a handful of boundary states 
at these points.  Furthermore their geometric
or non-geometric nature is obscure.
Their charges in a large radius basis are computable
following the techniques in \cite{BDLR:Dq}; but the moduli
space or any geometric interpretation of these objects
as submanifolds or otherwise is difficult to find.

In fact, we can do better by studying general aspects of the
open string worldsheet.
Suppose our target space is a \CY\ variety $X$,
and we study D-branes which preserve an $N=2$
worldsheet supersymmetry, {\em i.e.}\ the
``B-type'' branes of \cite{OOY:Dm}.
It was first proposed by Kontsevich \cite{Kon:mir} and then more
recently by Douglas \cite{Doug:DC} that (at least in the context of
zero string coupling) B-type D-branes are actually
objects in the bounded derived category of coherent sheaves $\DC(X)$.
Further work related to forging a link between D-branes and $\DC(X)$
has also appeared in \cite{Shrp:DC,LMW:McK,lsm:multistep} for example.
In this paper we will analyze this connection in detail and assess to
what extent this is true.  Using this description we will begin to
see how classical geometric notions of a D-brane can break
down when stringy geometry becomes important.

The key idea will be to define a specific 
set of topological field theories
with target space $X$. The sum of all such topological field theories
is then equivalent to $\DC(X)$. 
The objects in $\DC(X)$ can be called
``topological branes'' following \cite{Doug:DC}.
Building these topological field theories is a
three-step process:
\begin{enumerate}
\item Begin with Witten's ``B-model'' of \cite{W:AB,W:CS}.
\item Add the notion of integral ``grading'' following the
ideas of Douglas \cite{Doug:DC}. 
\item Consider specific kinds of ``marginal'' deformations of this
theory to form a more general class.
\end{enumerate}
Defining this set of topological quantum field theories and showing
it forms $\DC(X)$ constitutes section \ref{s:DX} of this paper.%
This section borrows heavily from ideas by Kontsevich
\cite{Kon:mir,Kont:mon} and especially Douglas 
\cite{Doug:DC} but some of the
details of our construction are a little different from previous work
and we try to spell out each step fairly explicitly given the complexity
of the subject.
Note that we address all of the words ``bounded'', ``derived
category'' and ``coherent'' in ``bounded derived category
of coherent sheaves''!

Such topological field theories can often be associated to twisting
the $N=2$ CFT associated to a configuration of branes.  When the
volume of $X$ is large, a single D-brane wrapped around a holomorphic
cycle is expected to preserve target space supersymmetry and therefore
B-type D-branes should be topological branes in the above sense, {\em
i.e.}\ objects in $\DC(X)$.  On the other hand, the converse statement
that all objects in $\DC(X)$ correspond to physical D-branes is far
from true.

The great advantage of using the derived category picture of D-branes
as opposed to, say, K-theory is that the former
picture contains considerably more data
specifying the D-brane. For example, 0-branes
at different points in $X$
are different objects in $\DC(X)$.
We will use this extra knowledge to analyze
the stability conditions on 0-branes in a \CY\ threefold. In section
\ref{s:D0} we will see
that for any \CY\ threefold embedded as a complete intersection in a
toric variety, one may make any given 0-brane unstable by a process
involving shrinking the overall size of the \CY.

In section \ref{s:disc} we briefly discuss three issues. First,
we discuss the relation of the category of topological
branes to the stable branes at a given point in the
K\"ahler moduli space, and make some suggestions
about the origins of monodromy.  We will then
discuss briefly the appearance in the open string
linear sigma models of \cite{lsm:monad,lsm:multistep,GJ:lsmseq}
of structures similar to those found in this 
paper and in \cite{Doug:DC}.
Finally, we discuss a very easy proof showing how $\DC(X)$ is
often invariant under a birational transformation of $X$ taking us to
another \CY.


\section{The Derived Category and Topological D-Branes}  \label{s:DX}


The purpose of this section is to give a complete proof that the sum
of a certain class of topological field theories on an algebraic
variety $X$ is equivalent (in a sense to be made precise) to the
bounded derived category of coherent 
sheaves. Many of the ideas of this section are copied from Douglas
\cite{Doug:DC}. Some of the details of the model are a little
different from Douglas and the logical order of the proof is
changed. We hope that this section clarifies some of the aspects of
the technically difficult subject of associating D-branes to objects
in the derived category.

\subsection{Witten's topological field theory}
                \label{ss:wtqft}

We begin with a finite-dimensional complex vector bundle $E$ with
connection $A$ over a complex \CY\ manifold
$X$. Imagine open strings propagating in $X$,
with Chan-Paton factors living in $E$.
Ref. \cite{W:CS} introduced an associated 
topological ``B-model'' for
maps $\Phi$ of a worldsheet $\Sigma$, with boundaries $C_k$, into $X$. 
The action for the bulk of the worldsheet is:
\begin{equation}
  \int_{\Sigma} \Bigl( g_{IJ}\partial_z\Phi^I\partial_{\bar z}\Phi^J +
  i\eta^{\bar\imath}(D_z\rho_{\bar z}^i + D_{\bar z}\rho_{z}^i)
  g_{i\bar\imath} + i\theta_i(D_{\bar z}\rho_{z}^i-
  D_z\rho_{\bar z}^i) +
  R_{i\bar\imath j\bar\jmath}\rho^i_z
  \rho^j_{\bar z}\eta^{\bar\imath}\theta_kg^{k\bar\jmath}\Bigr)d^2z\ ,
\end{equation}
and is identical to the action for the topological ``B-model''
defined on closed string worldsheets \cite{W:AB}.
Here $\theta$ and $\eta$ are fermions transforming 
as 0-forms on $\Sigma$; 
and $\rho$ is a fermion transforming as a 1-form. 
The indices $I,J$ represent
real coordinates on $X$ and the corresponding 
lower-case indices represent
(anti)holomorphic coordinates on $X$.

For each boundary $C_k$ of $\Sigma$,
there is an additional term in the action:
\begin{equation}
  S_{C_k} = \oint_{C_k} \Bigl(\Phi^*(A) - i\eta^{\bar\imath}
   F_{\bar\imath j}\rho^j\Bigr)\ .
\end{equation}
This appears in the path integral with a trace over the
structure group of $E$:
\begin{equation}
	\Tr_k {\rm P} \exp\left(- S_{C_k}\right)\ .
\end{equation}

This topological quantum field theory on $\Sigma$ has a BRST symmetry
whose transformation laws are given by
\begin{equation}
\begin{split}
  \delta\Phi^i &= 0\\
  \delta\Phi^{\bar i} &= i\alpha\eta^{\bar\imath}\\
  \delta\eta^{\bar\imath} &= 0\\
  \delta\theta_i &= 0\\
  \delta\rho^i &= -\alpha\,d\Phi^i\ ,
\label{eq:brsttrans}
\end{split}
\end{equation}
where $\alpha$ is a globally-defined
fermionic parameter. The BRST operator $Q$ is then
defined by $\delta\Lambda = -i\alpha\{Q,\Lambda\}$ for any field
$\Lambda$.  The invariance of the action 
under $Q$ requires that $A$ be a {\em
holomorphic\/} connection. That is, the (2,0) and (0,2) part of the
curvature $F_{i\bar\jmath}$ must vanish.  Thus the open string
``B-model'' exists for ``B-type'' boundary conditions \cite{BDLR:Dq}
as defined in \cite{BBS:5b,OOY:Dm}.

The operators in this topological field theory are then given by BRST
cohomology. For open string states the operators correspond to the
bundle-valued {\em Dolbeault\/} cohomology:
\begin{equation}
\phi\in H^{0,p}(X,\End E)\ .
\end{equation}
The ghost number of a given operator is given by $p$.
The operator product algebra is then generated by this cohomology with
the product being given by the usual wedge product together with the
natural composition for the group $\End E$.

For closed string states some of the BRST cohomology
corresponds to deformations
of the complex structure of the $X$.  Deformations
of the {\em K\"ahler}\ structure are BRST-exact
and decouple from the theory \cite{W:AB,Doug:DC,BDLR:Dq}.\footnote{The
proof in \cite{BDLR:Dq} of this statement for open strings
showed decoupling
only up to a boundary term.  This term is canceled
by a boundary contact term added to
the integrated vertex operator, as in \cite{Gut:cntc}.}
Therefore the open topological B-model is independent
of the K\"ahler deformations of the theory.

Finally, the correlators in the topological theory receive
contributions only from constant maps into the target space.
Furthermore, there is a ``ghost number'' (fermion number)
selection rule such that correlators on the disc must violate
ghost number by exactly the dimension of $X$ to be nonvanishing.


\subsection{Adding the grading}  \label{ss:grade}

So far we have discussed the topological sigma model associated
to $N$ D-branes which wrap the entire \CY\ manifold $X$,
and some vector bundle on them. Here $N$ is the rank of bundle. In other
words, for a \CY\ threefold we have just discussed $N$
D6-branes.\footnote{Our notation is that a D$p$-brane wraps $p$
real dimensions of $X$. That is, if we compactify on $X$, a D$p$-brane
appears as a particle in the non-compact dimensions.}
We consider this system a ``single'' D-brane.
We wish to generalize this to include strings stretching
between a finite collection of such objects.  
Eventually we will consider deformations of the topological field theory
which encompass much more general situations.

One of the most important observations by Douglas in \cite{Doug:DC} is
the essential r\^ole of a {\em grading\/} for D-branes. Consider the
(untwisted) $N=2$ worldsheet theory for strings. In the case of closed
strings there is a left-moving and a right-moving spectral flow
operator for the internal CFT, 
both of which make up part of the operator generating
spacetime supersymmetry transformations. When one
passes to open strings, the left-moving and right-moving sectors mix
on boundaries and these two spectral flow operators cease to be
independent.  If the D-brane at one
end of the string preserves half
of the spacetime supersymmetries, 
the linear combination of spectral flow
operators which vanishes at the boundary corresponds
to the unbroken spacetime supercharge.

To each D-brane we can associate 
the phase shift between the left-moving and right-moving
spectral flow operator at the boundary ending on that
D-brane.  This phase is the ``grade'' of the D-brane.
Since the spectral flow operator can be constructed from
the bosonized $\GU(1)_R$ current of the $N=2$ worldsheet
algebra, the grade lives in $\GU(1)$. 
Two branes are mutually supersymmetric 
only if their grade coincides. At this
point the topological operators will
have integral $U(1)$ charge as demanded
by spacetime supersymmetry.

The mirror of our B-type D-branes
gives an intuitive picture of the relative
grading. For A-type branes associated to special Lagrangian submanifolds,
the relative grading of two branes
is literally the angle between them. Whether
or not spacetime supersymmetry
is broken can be determined by this angle
\cite{BDL:angles,DFR:stab}. Within this
framework, the strings stretching between
the two D-branes can be created by a ``twist field''
changing the moding of the worldsheet fields;
the grade is simply the worldsheet
$\GU(1)$ charge of this twist field \cite{BDL:angles}.

More precisely, the grade for an A-brane wrapped on a
special Lagrangian submanifold $\Gamma$ is the phase of the 
associated period:
\begin{equation}
  \frac1\pi\Img\log\int_\Gamma \Omega\ .  \label{eq:Agrade}
\end{equation}
The normalization is fixed so that the grading is initially defined
mod 2.  If the A-brane corresponds to a BPS particle
in four dimensions, this grading is just the
phase of the central charge of the brane \cite{DFR:stab}.

If we vary the complex structure of the target space, this grading for
A-branes will vary. In particular if we initially define the relative
grading of the ends of a string to be in the range $[0,2)$, we can
easily move in moduli space so that the relative grading lies outside
this range. Because of this we allow the grading to be defined in $\R$
rather than $\GU(1)$. Note that there is no absolute meaning
to a shift of the grading by two; such a shift should therefore
considered a gauge symmetry.

Shifting the grade by 1, but otherwise preserving
the boundary conditions, changes the linear
combination of $N=2$ spacetime supercharges
that the D-brane preserves.  This adds a factor of $(-1)$ to
the central charge, indicating that the shifted brane
is simply the anti-brane.  We may therefore refine the
gauge symmetry of the previous paragraph
by defining it to be {\em a shift in the grading by one,
combined with exchanging the r\^ole of branes and anti-branes.}

In the open string CFT, the grading for B-branes is determined by
$B+iJ$ and is given essentially by the mirror of (\ref{eq:Agrade}):
\begin{equation}
  \frac1\pi\Img\log\int_X e^{B+iJ}\ch(E)\sqrt{\td(T_X)}+\ldots,  
	\label{eq:Bgrade}
\end{equation}
where $E$ is the bundle (or sheaf or 
complex of sheaves etc.) representing the D-brane. 

This grading depends essentially on $B+iJ$; therefore
it will not appear as physical data in the topological
B-model.  Nonetheless, 
we wish to modify Witten's B-model to include some notion of
grading. We will do this by decomposing our collection of
branes as:
\begin{equation}
  E = \bigoplus_{n=-\infty}^\infty E^n\ .
\label{eq:decompose}
\end{equation}
Since $E$ is finite dimensional, only a finite number of the bundles
$E^n$ are nonzero. The index
$n\in\Z$ defines the grading associated to the bundle $E^n$.
The end of a given open string must be associated to a definite
grading and hence a single summand $E^n$.

It is important to note two essential differences between the gradings
for the physical B-branes constructed in the untwisted CFT, and the
grading we have introduced into our topological field theory:
\begin{enumerate}
\item The gradings in the topological field theory have been fixed to be
integers. In this sense the topological field theory is {\em less\/} general
than the physical B-branes.
\item The gradings are fixed and do not depend upon $B+iJ$. There may
be no value of $B+iJ$ for the physical branes which yields a given
choice of gradings in the topological field theory. In this sense the 
topological field theory is {\em more\/} general
than the physical B-branes.
\end{enumerate}
Note also that since the gradings do not depend on $B+iJ$, our
modified topological field theory is still invariant under
deformations of the K\"ahler structure.

There is still a strong connection between our topological field theory and
physical B-branes. Suppose we consider a combination of branes
and anti-branes which preserve spacetime supersymmetry.%
\footnote{We call an
anti-brane relative to a brane an object whose grading
differs by an odd integer from the brane.}  By an overall
shift of the grading (which is physically meaningless) we can make the
gradings of all the branes an even integer and the gradings of all the
anti-branes an odd integer. If we topologically
twist this theory following \cite{W:AB,W:CS,EY:twist}
we obtain a
topological field theory consistent with our assumptions above. 
Following \cite{Doug:DC} we will refer to the boundaries in our
topological field theory as ``topological D-branes'' to distinguish
them from physical D-branes.
We will address the connection between physical
topological branes further in section \ref{ss:phys}.
We note for now, however, that any single D-brane by itself preserves
supersymmetry and so falls into the class of interest.

When adding the notion of grading, very little is changed in Witten's
B-model. Operators are now elements 
of:
\begin{equation}
  \phi\in H^{0,p}(X,(E^m)^\vee\otimes E^n)\ .  \label{eq:phi1}
\end{equation}
Note that the open strings are {\em oriented\/}. The above string goes
from $E^m$ to $E^n$.
The grading is associated to the $\GU(1)_R$ symmetry of the $N=2$
worldsheet theory. 
To maintain the relationship to the untwisted theory, we
define our topological field theory so that 
grading is associated to the ghost number. For
field $\phi$ in (\ref{eq:phi1}), this is given by $p+n-m$.

It is now useful to change language a little. Rather than thinking of
$E^n$ as a complex vector bundle over the complex manifold $X$, we
consider $\cE^n$ to be the 
associated {\em locally free sheaf\/} over
the algebraic variety $X$. Using ideas of sheaf cohomology we can
rewrite (\ref{eq:phi1}) as (see section III.6 of \cite{Hartshorne:}
for example)
\begin{equation}
  \phi\in \Ext^p(\cE^m, \cE^n)\ .  \label{eq:phi2}
\end{equation}
The product of operators is given by the ``Yoneda pairing''
\begin{equation}
  \Ext^p(\cE^l,\cE^m)\otimes\Ext^q(\cE^m,\cE^n)\to\Ext^{p+q}(\cE^l,\cE^n)\ ,
        \label{eq:prod1}
\end{equation}
signifying two open strings joining along a common boundary $\cE^m$.


\subsection{A category}   \label{ss:cat}

In this section we will describe the topological field theory of the
previous sections as a {\em category}. At first sight this looks like
introducing mathematical mumbo jumbo without any real need for it. We
hope that by the end of this paper the reader will be convinced that
category theory is indispensable for describing D-branes. We would
also like to point out that this is not the same use of category
theory that was used in the context of topological field theory in
knot theory (as in \cite{Atiyah:knotbook} for example). The category
language for D-branes has also been used in other works such as
\cite{Laz:Dcat}. 

The objects in our category will be the finite collection of
nontrivial sheaves $\cE^n$ and the morphisms will be the operators
$\phi$ from (\ref{eq:phi2}). In other words, the objects are D-branes
and the morphisms are open strings. Composition of morphisms is then
given by the Yoneda pairing (\ref{eq:prod1}).

There are only two conditions that these objects and morphisms need to
satisfy in order for this to be a category:
\begin{itemize}
\item For every object there exists an identity morphism. Clearly
this is given by $1_n\in\Ext^0(\cE^n,\cE^n)$.
\item The morphisms are associative. This is equivalent to the
associativity of the operator product algebra of the topological field
theory.\footnote{The reader may recall that in string {\em field\/}
theory the multiplication of states is not always
associative \cite{HS:assoc}.
This arises in considering open string states created by the
open-closed string vertex.  The multiplication of 
open string vertex operators is associative, however.}
\end{itemize}

The content of a topological field theory is given completely by its
operator algebra. The topological field theory we are discussing is
therefore completely equivalent to this category.

Note that by construction we have described a category with a finite
number of objects. Later on we will also consider the category
$\CT(X)$ of all possible topological D-branes on $X$
\def\DC{\mathbf{D}}. This latter category has an infinite number of
objects. The category associated to a particular topological field
theory is a full subcategory of $\CT(X)$.


\subsection{Deformations}  \label{ss:def}

So far we do not have anything that resembles the derived category of
coherent sheaves. The topological field theory we have discussed so
far is not general enough. In this section we will look for
deformations of Witten's B-model.

A deformation of a topological field theory amounts to adding a
Q-closed object to the action. More specifically,
in a topological CFT,
take a Q-closed ``local'' boundary operator $\varphi$ with
ghost charge $h$.  An integrated boundary vertex operator,
suitable for adding to the action, is \cite{W:AB,EY:twist,BDLR:Dq}:
\begin{equation}
	\delta_\varphi S = t \oint_{C_k} \left\{ G, \varphi\right\}\ .
\end{equation}
where $G$ is the fermionic spin-one current
in the twisted $N=2$ superalgebra.  The BRST variation
of the integrand is a total derivative.  This deformation
has ghost number $(h - 1)$.

Adding $\delta_\varphi S$ to the topological worldsheet action
implies adding $\varphi$ as a boundary term
to the BRST current $Q_0$.  To see
this we can vary the integrand according to (\ref{eq:brsttrans})
with $\alpha$ having arbitrary dependence on the 
boundary coordinate $\tau$.  The coefficient of 
$\partial_\tau \alpha$ in $\delta S$ is the deformation 
$\delta Q$ of the BRST charge.  
Since $\left\{G, Q\right\} = \partial_\tau$, after an
integration by parts we find that 
\begin{equation}
	\delta Q = t \varphi \equiv d\ ,
\end{equation}
where $\varphi$ is supported on the boundary $C_k$.

In order that
the topological field theory retain its possible identity as a twisted version
of an $N=2$ SCFT we require that $\varphi$ has ghost number
one. This is equivalent to demanding that it appears as a marginal
operator in the untwisted theory. The candidate ghost number one
operators are
\begin{equation}
\begin{split}
  \Ext^0&(\cE^n,\cE^{n+1})=\Hom(\cE^n,\cE^{n+1})\\
  \Ext^1&(\cE^n,\cE^n)\\
  \Ext^2&(\cE^n,\cE^{n-1})\\
  &\quad\vdots
\end{split}
\end{equation}
and so on up to the dimension of $X$.

The group $\Ext^1(\cE^n,\cE^n)$ represents first-order deformations of the
sheaf $\cE^n$.  These yield obvious deformations of
Witten's B-model.

Of considerably more interest are the operators living in
$\Hom(\cE^n,\cE^{n+1})$. These will be of central importance to us in
this paper. The higher operators $\Ext^2(\cE^n,\cE^{n-1})$, etc.\ will
produce yet more deformations of the topological field theory. While
these deformations will be distinct from $\Ext^1(\cE^n,\cE^n)$ and
$\Hom(\cE^n,\cE^{n+1})$, we will see in section \ref{ss:tachy} that,
once we introduce the derived category, they essentially add nothing new.

Our more general topological field theory will therefore be described
by a set of locally free sheaves $\cE^n$ together with holomorphic maps
(i.e., morphisms in the category of sheaves)
\begin{equation}
  d_n:\cE^n\to \cE^{n+1}\ .
\end{equation}
which correspond to marginal operators in the topological
sigma model.  As these operators map between
different D-branes, they are ``boundary condition-changing
operators'' in the sense of \cite{Cardy:bc}.

Upon turning on these deformations, we have argued that
$Q$ becomes $Q_0+d(\sigma=0)+d(\sigma=\pi)$,
where $Q_0$ is the original BRST operator of section \ref{ss:wtqft}
and $d$ on each boundary is a sum of the associated
ghost number 1 operators $\varphi$.
In order that the deformed theory remain
topological, the deformations must be integrable;
\ie\ they must be invariant under the deformed
BRST charge.    This implies:
\begin{equation}
  d_{n+1}d_n = 0\ , \qquad\forall n\ ,
\end{equation}
which is also necessary for the nilpotency of $Q_0 + \delta Q$.
In other words, the topological field theory is now specified by a
{\em complex\/} of locally free sheaves:
\begin{equation}
\xymatrix@1{
\ldots\ar[r]&\cE^{-1}\ar[r]^{d_{-1}}&\cE^0\ar[r]^{d_0}
&\cE^1\ar[r]^{d_1}&\ldots
}
\end{equation}
Since the original vector bundle $E$ in section \ref{ss:wtqft} was
finite-dimensional, this complex is {\em bounded}.\footnote{All
complexes and derived categories in this paper are bounded from now on
whether we explicitly state this or not.}
We will denote this complex $\cE^\bullet$.


\subsection{The new $Q$-cohomology}  \label{ss:Qcoh}

Now that we have deformed our topological field theory and thus the
$Q$-operator, we need to recompute the $Q$-cohomology to find the
operator algebra of the new topological field theory. We will discover
that the desired cohomology is ``{\em hyperext\/}'' (see section
10.7 of \cite{Wei:hom} for example). That is, if you were a
mathematician familiar with dealing with the homological algebra of
complexes, $Q$-cohomology is precisely what you would hope it is!

First note that we are no longer free to associate the end of the
string with a particular $\cE^n$.
We must include both boundary operators defined for a given
boundary condition, and boundary condition-changing operators.
In other words, switching on the $d$-maps generically
tangles the terms in the complex together into one D-brane.

In this section we would like to consider a string stretching between
two possibly {\em distinct\/} D-branes. We may achieve this as follows.

Assume that each $\cE^n$ is actually a direct sum of two sheaves for
all $n$. Without trying to confuse the reader too much we will call
this sum $\cE^n\oplus \cF^n$.  Now we will restrict the maps in this
complex to being block-diagonal. In other words we have maps
\begin{equation}
\begin{split}
d^E_n&:\cE^n\to \cE^{n+1}\\
d^F_n&:\cF^n\to \cF^{n+1}\ ,
\end{split}
\end{equation}
with no $d$ maps mixing the $\cE$'s and $\cF$'s. 

This gives us the notion of two D-branes --- one associated to
 $\cE^\bullet$ and one associated to $\cF^\bullet$.
We can then assume that the boundary conditions for the start of the
string are given by $\cE^\bullet$ and the end of the string are given by
$\cF^\bullet$.

In order to compute how $Q$ acts on such a string we will need to make
use of the idea of collapsing a double complex into a single complex.
Let us first consider the {\em sheaf\/}\footnote{Do not confuse this
with the vector space $\Hom(\cE^m,\cF^n)$. $\sHom(\cE^m,\cF^n)$ is the sheaf
which associates an open set $U$ with local holomorphic maps from
sections of the bundle $E^m$ over $U$ to sections of the the bundle
$F^n$ over $U$.} given by $\sHom(\cE^m,\cF^n)$. The maps $d^E$ and
$d^F$ induce a double complex:
\begin{equation}
\xymatrix{
&\ar[d]^-{d^E_1}&\ar[d]^-{d^E_1}&\\
\ar[r]^-{d^F_{-1}}&\sHom(\cE^1,\cF^0)\ar[r]^{d^F_0}\ar[d]^{d^E_0}
& \sHom(\cE^1,\cF^1)\ar[r]^-{d^F_1}\ar[d]^{d^E_0} &\\
\ar[r]^-{d^F_{-1}}&\sHom(\cE^0,\cF^0)\ar[r]^{d^F_0}\ar[d]^-{d^E_{-1}}
& \sHom(\cE^0,\cF^1)\ar[r]^-{d^F_1}\ar[d]^{d^E_{-1}} &\\
&&&
}
\end{equation}
We may now form the single complex
\begin{equation}
\xymatrix{
  \ldots\ar[r]&\sHom^0(\cE^\bullet,\cF^\bullet)\ar[r]^{\bar d_0}
  &\sHom^1(\cE^\bullet,\cF^\bullet)\ar[r]^-{\bar d_1}&\ldots
}       \label{eq:Homc}
\end{equation}
by defining:
\begin{equation}
  \sHom^q(\cE^\bullet,\cF^\bullet) = 
\bigoplus_n\sHom(\cE^n,\cF^{n+q})\ ;
\end{equation}
and $\bar d=d^E+d^F$, where $d^E$ and $d^F$ anti-commute.

The cohomology of the complex (\ref{eq:Homc}) may be computed by using
a {\em spectral sequence\/} (see \cite{BT:} for example). In this case
we have a spectral sequence which is a mixture of the usual
homological and cohomological spectral sequence. 
One way to write it is that
\begin{equation}
  (E_0)^n_m = \sHom(\cE^m,\cF^n) \Rightarrow
        H^{n-m}(\sHom^\bullet(\cE^\bullet,\cF^\bullet))\ . \label{eq:ss1}
\end{equation}

Now the problem at hand is to compute the cohomology of $Q=Q_0+\bar
d$. Note that $Q_0$ and $\bar d$ anti-commute. This suggests
another double complex:
\begin{equation}
\xymatrix{
&&&\\
\ar[r]^-{\bar d}&\Omega^1(\sHom^0(\cE^\bullet,\cF^\bullet))
\ar[r]^{\bar d}\ar[u]^{Q_0}
& \Omega^1(\sHom^1(\cE^\bullet,\cF^\bullet))\ar[r]^-{\bar d}\ar[u]^{Q_0} &\\
\ar[r]^-{\bar d}&\Omega^0(\sHom^0(\cE^\bullet,\cF^\bullet))
\ar[r]^{\bar d}\ar[u]^{Q_0}
& \Omega^0(\sHom^1(\cE^\bullet,\cF^\bullet))\ar[r]^-{\bar d}\ar[u]^{Q_0} &\\
&\ar[u]^-{Q_0}&\ar[u]^-{Q_0}&
}   \label{eq:dcplxbig}
\end{equation}

Here we use the notation $\Omega^p$ for the complex of ``things'' for
which $Q_0$ is a boundary map. What is this exactly? We saw in section
\ref{ss:wtqft} that $Q_0$ basically acts as the boundary operator on the
twisted Dolbeault complex. That is, if we go back to differential
geometry, $\Omega^p(\sHom^q(\cE^\bullet,\cF^\bullet))$ may be thought
of as ``$(0,p)$-forms with values in the bundle associated to
$\sHom^q(\cE^\bullet,\cF^\bullet)$.''  We want to use the
corresponding sheaf cohomology for the vertical maps. That is, we
should do something along the lines of having injective resolutions of
$\sHom^q(\cE^\bullet,\cF^\bullet)$ in the vertical direction.

Anyway, consider a spectral sequence for the complex
(\ref{eq:dcplxbig}). Take cohomology in the horizontal direction and
then the vertical direction to obtain:
\begin{equation}
  E_2^{p,q} = H^p(X,H^q(\sHom^\bullet(\cE^\bullet,\cF^\bullet)))
  \Rightarrow H_Q^{p+q}\ ,   \label{eq:ltog}
\end{equation}
where $H_Q$ is the desired $Q$-cohomology.
Now (\ref{eq:ltog}) is a kind of ``local to global'' spectral
sequence (see section 4.2 of \cite{Grot:qp}). This implies that the
cohomology group 
$H_Q^P$ is actually 
given by the group $\Hom^P(\cE^\bullet,\cF^\bullet)$. We may describe
this group as follows. 

Any locally free sheaf $\cF^n$ has an injective resolution:
\begin{equation}
  0\to\cF^n\to I^0(\cF^n)\to I^1(\cF^n) \to\ldots,
\end{equation}
where $I^s(\cF^n)$ is an ``injective object'' in the category of
quasi-coherent sheaves. Note that such injective
objects are very peculiar things and look nothing like locally free
sheaves. We may then similarly replace the entire complex $\cF^\bullet$ by a
complex of injective objects. That is, replace $\cF^n$ in the complex
by $\bigoplus_s I^s(\cF^{n-s})$. The maps between $\cF^n$ then have
natural lifts to the maps between $I^s(\cF^{n-s})$. We will denote
this complex of injectives $\cF^\bullet_{\mathrm{inj}}$.

Now define a complex $\hom^\bullet(\cE^\bullet,\cF^\bullet)$ by
\begin{equation}
  \hom^P(\cE^\bullet,\cF^\bullet) = \bigoplus_n
  \Hom(\cE^n,\cF^{n+P}_{\mathrm{inj}})\ , 
\end{equation}
and the obvious maps
$\hom^P(\cE^\bullet,\cF^\bullet_{\mathrm{inj}})\to
\hom^{P+1}(\cE^\bullet,\cF^\bullet_{\mathrm{inj}})$ 
induced by $d^E+d^F$.
One may show that the groups $\Hom^P(\cE^\bullet,\cF^\bullet)$ are
then given by the cohomology of this chain complex.

We have therefore defined the group $\Hom^P(\cE^\bullet,\cF^\bullet)$
in which the operators of the
topological field theory live. Clearly $P$ is the ghost number. The
group $\Hom^P(\cE^\bullet,\cF^\bullet)$ is often referred to as
``hyperext'' in the mathematical literature and might also be denoted
$\Ext^P(\cE^\bullet,\cF^\bullet)$.

The operator product is a simple generalization of the Yoneda pairing
\begin{equation}
\Hom^P(\cE^\bullet,\cF^\bullet)\otimes
\Hom^Q(\cF^\bullet,\cG^\bullet)\to
\Hom^{P+Q}(\cE^\bullet,\cG^\bullet)\ .
\end{equation}


\subsection{Enter the derived category}  \label{ss:DC}

Let $\CT(X)$ be the category of all possible topological field theories
of the type considered in section \ref{ss:Qcoh} with target variety
$X$. We need to know precisely what we mean by this.
Physically we desire that the objects of $\CT(X)$ form all
possible D-branes of 
the type we are considering and the morphisms are open strings.

From the above analysis one would be tempted first to assert that
the objects of $\CT(X)$ are all possible bounded complexes of
locally free sheaves and the morphisms are the open string operators
given by $\Hom^P(\cE^\bullet,\cF^\bullet)$. This is not quite right
however since it distinguishes between physically identical D-branes.

We need to consider
more carefully when two different complexes really should be
considered ``different'' objects in $\CT(X)$. In order to obtain the
right mathematical description of $\CT(X)$ we will need to divide the
category of chain complexes by physical equivalences. This leads to
the derived category.

The physical content of a topological field theory is completely
described by the operator product algebra. Let us consider two
D-branes described by complexes of sheaves $\cE^\bullet_1$ and
$\cE^\bullet_2$. These are physically identical D-branes if and only
if
\begin{equation}
\begin{split}
\Hom^P(\cE^\bullet_1,\cF^\bullet) &= \Hom^P(\cE^\bullet_2,\cF^\bullet)
	\quad\textrm{and}\\
\Hom^P(\cF^\bullet,\cE^\bullet_1) &= \Hom^P(\cF^\bullet,\cE^\bullet_2)
\end{split}
\end{equation}
for all $P$ and all D-branes $\cF^\bullet$.
We therefore construct $\CT(X)$ as the category of complexes and
morphisms divided by this equivalence relationship.

We will show in this section that
$\CT(X)$ is very closely related to $\DC(X)$, the derived category of
coherent sheaves in $X$. 
This requires only standard manipulations in homological
algebra. Indeed, any reader familiar with the derived category and
hyperext will probably find the following unnecessarily verbose.
We will use the definition/theorem of
the derived category $\DC(\mathbf{A})$ of a given category $\mathbf{A}$
from chapter 4, 1.3 of \cite{GM:Hom}:
\begin{theorem}
Let $\mathbf{A}$ be an abelian category, 
and let $\mathbf{Kom}(\mathbf{A})$ be
the category of complexes over $\mathbf{A}$. There exists a category
$\DC(\mathbf{A})$ and a functor $\mathsf{Q}:\mathbf{Kom}(\mathbf{A})\to
\DC(\mathbf{A})$ with the following properties:
\begin{enumerate}
\item $\mathsf{Q}(f)$ is an isomorphism for any quasi-isomorphism $f$.
\item Any functor $\mathsf{F}:\mathbf{Kom}(\mathbf{A})\to\mathbf{C}$
transforming quasi-isomorphisms into isomorphisms can be uniquely
factorized through $\DC(\mathbf{A})$, i.e., there exists a unique
functor $\mathsf{G}:\DC(\mathbf{A})\to\mathbf{C}$ with $\mathsf{F}=
\mathsf{G}\circ\mathsf{Q}$.
\end{enumerate}  \label{th:1}
\end{theorem}
A ``quasi-isomorphism'' is a chain map which induces an
isomorphism on the cohomology groups of the complex.

\def\Kloc{\mathbf{K}_{\mathrm{LF}}}
Let $\Kloc(X)$ be the category of complexes of locally free sheaves on
$X$. The morphisms in this category are chain maps. We wish to
construct a functor $\mathsf{F}:\Kloc(X)\to\CT(X)$. The definition of
this functor is pretty obvious. A chain in $\Kloc(X)$ maps to the
corresponding D-brane in $\CT(X)$. A chain map in $\Kloc(X)$ maps to
an element of $\Hom^0(\cE^\bullet,\cF^\bullet)$ in $\CT(X)$. Note that
homotopic chain maps are identified by $\mathsf{F}$.

Now we wish to argue that a quasi-isomorphism
in $\Kloc(X)$ maps to an isomorphism in $\CT(X)$. Note that the
cohomology of a complex of locally-free sheaves is a set of {\em
coherent sheaves\/} in general but this still allows us to define the
notion of a quasi-isomorphism in $\Kloc(X)$.

Consider a quasi-isomorphism between two complexes
$f:\cE^\bullet_1\to\cE^\bullet_2$. Clearly this induces a map
\begin{equation}
  f^*:\Hom^P(\cE^\bullet_2,\cF^\bullet)\to\Hom^P(\cE^\bullet_1,
	\cF^\bullet)\ ,
        \label{eq:fstar}
\end{equation}
from the diagram (for e.g., $P=1$)
\begin{equation}
\xymatrix{
\ar[r]&\cE^0_1\ar[r]\ar[d]^f&\cE^1_1\ar[r]\ar[d]^f&\cE^2_1\ar[r]\ar[d]^f&\\
\ar[r]\ar[rd]&\cE^0_2\ar[r]\ar[rd]&\cE^1_2\ar[r]\ar[rd]&\cE^2_2\ar[r]\ar[rd]&\\
\ar[r]&\cF^0\ar[r]&\cF^1\ar[r]&\cF^2\ar[r]&
}
\end{equation}
We would like to show that $f^*$ is actually an isomorphism.

Consider first the case where $\cE^\bullet$ is acyclic, i.e., a complex
with trivial cohomology and let $\cF^\bullet$ be any complex. The spectral
sequence (\ref{eq:ss1}) tells us that
the cohomology of $\sHom^\bullet(\cE^\bullet,\cF^\bullet)$ is
trivial. The spectral sequence (\ref{eq:ltog}) then tells us that the
$Q$-cohomology of the topological field theory is trivial. Similarly
if $\cF^\bullet$ is acyclic then $\Hom^P(\cE^\bullet,\cF^\bullet)=0$
for any $\cE^\bullet$.

Now we need to introduce the ``Cone'' of
a map $f$ of complexes. 
The mapping cone $\Cone(f:\cE^\bullet\to\cF^\bullet)$ is
defined as the complex 
\begin{equation}
\xymatrix@1@C=15mm{
\ar[r]&\cE^1\oplus\cF^0\ar[r]^{\left(\begin{smallmatrix}d_E&0\\
f&d_F \end{smallmatrix}\right)}&
\cE^2\oplus\cF^1\ar[r]^-{\left(\begin{smallmatrix}d_E&0\\
f&d_F \end{smallmatrix}\right)}&
\cE^3\oplus\cF^2\ar[r]&\ldots
}  \label{eq:cone}
\end{equation}
This mapping cone has the very useful property that
$\Cone(f:\cE^\bullet\to\cF^\bullet)$ is acyclic if and only if the
map $f$ is a quasi-isomophism (see section 1.5 of \cite{Wei:hom} for
example).

Now return to the quasi-isomorphism
$f:\cE^\bullet_1\to\cE^\bullet_2$. 
The map $f$ will induce a chain map $f^\sharp$ as follows:
\begin{equation}
\xymatrix{
\ar[r]&\hom^0(\cE_1^\bullet,\cF^\bullet)\ar[r]
  &\hom^1(\cE_1^\bullet,\cF^\bullet)\ar[r]&\\
\ar[r]&\hom^0(\cE_2^\bullet,\cF^\bullet)\ar[r]\ar[u]_{f^\sharp}
  &\hom^1(\cE_2^\bullet,\cF^\bullet)\ar[r]\ar[u]_{f^\sharp}&
}
\end{equation}
If $f$ is a quasi-isomorphism then its cone is acyclic. Thus
the groups $\Hom^P(\Cone(f),\cF^\bullet)$ associated to the cone
are zero. This in turn implies that the cone of $f^\sharp$ is acyclic
which in turn shows that $f^\sharp$ is a quasi-isomorphism. Thus
$\Hom^P(\cE^\bullet_1,\cF^\bullet)\cong\Hom^P(\cE^\bullet_2,\cF^\bullet)$
and (\ref{eq:fstar}) provides the canonical isomorphism.

Similarly one may show that for a quasi-isomorphism
$g:\cF^\bullet_1\to\cF^\bullet_2$, the induced map
$g_*:\Hom^P(\cE^\bullet,\cF^\bullet_1)\to\Hom^P(\cE^\bullet,\cF^\bullet_2)$
is an isomorphism.

We have shown that a quasi-isomorphism in $\Kloc(X)$ maps to an
isomorphism in $\CT(X)$. However, in order to use theorem~\ref{th:1} we
need to consider complexes over an {\em abelian\/} category. The
category of locally-free sheaves is not abelian, since it does not
contain its own cokernels. We may consider the larger category of {\em
coherent\/} sheaves instead. Let $\mathbf{Kom}(X)$ denote the category
of bounded complexes of coherent sheaves on $X$.

In order to define a functor from $\mathbf{Kom}(X)$ to $\CT(X)$ we
need to define the image of complexes of coherent sheaves which are
not locally free. This is actually very easy given our above analysis
of quasi-isomorphisms.

Given any coherent sheaf we can find a locally free resolution
and hence a complex of locally free sheaves which is quasi-isomorphic
to a complex containing only our original coherent sheaf. Furthermore
given any map between coherent sheaves we may find locally free
resolutions which allow this map to be lifted to a map between the
complexes of locally free sheaves. Therefore any configuration of
branes involving coherent sheaves may be rewritten in terms of locally
free sheaves.

All this shows that $\CT(X)$ already contains the image of complexes
of all coherent sheaves.
In a way we have justified various conjectures in the past
(e.g. \cite{HM:alg2,GS:sh}) that {\em coherent\/} sheaves can be relevant for
describing D-branes. 

We have now constructed a functor from $\mathbf{Kom}(X)$ to
$\CT(X)$. Therefore by theorem~\ref{th:1} we have constructed a
functor
\begin{equation}
\mathsf{G}:\DC(X)\to\CT(X)\ .  \label{eq:G}
\end{equation}


\subsection{$\mathsf{G}$ as an equivalence of categories}   \label{ss:func}

The statement that the category of all D-branes in our
topological field theories is the same as the derived category of
coherent sheaves on $X$ amounts to saying that the functor $\mathsf{G}$
in (\ref{eq:G}) is an equivalence of categories.

As it stands this is not quite true. However, we may make some fairly innocuous
changes to make it true. Let us define the category $\CT_0(X)$ which
is a subcategory of $\CT(X)$. $\CT_0(X)$ contains exactly the same
objects as $\CT(X)$ but we only consider morphisms
$\Hom^0(\cE^\bullet,\cF^\bullet)$. That is, we throw out all open
strings of nonzero ghost number. It is easy to argue that $\CT_0(X)$
is equivalent to $\DC(X)$ as we now show.

To show that $\mathsf{G}:\DC(X)\to\CT_0(X)$ is an equivalence of
categories we need to show that $\mathsf{G}$ is ``full, faithful, and
dense'' (see for example section 14 of \cite{HS:catg}).

The ``dense'' property asserts that every object in $\CT_0(X)$ is
isomorphic to some object in the image of $\mathsf{G}$. This is clear
--- every D-brane can be represented by a complex of coherent sheaves.

Now fix a pair of objects $\cE^\bullet$, $\cF^\bullet$ in
$\DC(X)$. The morphisms between $\cE^\bullet$ and $\cF^\bullet$ come
from chain maps up to various equivalence relations upon building
$\DC(X)$. These morphisms are given exactly by
$\Hom^0(\cE^\bullet,\cF^\bullet)$ --- the ghost number zero open
strings in $\CT_0(X)$. Thus $\mathsf{G}$ is an isomorphism for
morphisms between any pair $\cE^\bullet$, $\cF^\bullet$. This shows
that $\mathsf{G}$ is ``full'' and ``faithful''.

Having shown that $\mathsf{G}:\DC(X)\to\CT_0(X)$ is an equivalence of
categories, it is fairly trivial to see that $\CT_0(X)$ contains
essentially the same 
information as $\CT(X)$. It is easy to show that
\begin{equation}
\Hom^P(\cE^\bullet,\cF^\bullet)=
	\Hom^0(\cE^\bullet,\cF^\bullet[P])\ ,
\end{equation}
where $[P]$ means shift the complex $P$ places to the left. This means
that the open strings with nonzero ghost number are also found in
$\CT_0(X)$.

In this sense {\em the category of topological field theories given by
$\CT(X)$ is essentially the same thing as the bounded derived category
of coherent sheaves $\DC(X)$.}

\subsection{``Tachyon'' condensation}  \label{ss:tachy}

Finally in our review and analysis of the r\^ole of the derived category we
would like to discuss the ``tachyon condensation'' of \cite{Sen:dbd}.%
\footnote{Tachyon condensation and brane-anti-brane annihilation
in the topological A-model has been discussed in \cite{Vafa:bbar}.}
This is nothing more than giving a vacuum expectation value to a
string vertex operator in the worldsheet theory.  In the physical
theory the tachyon corresponds to a relevant boundary operator.
One can understand the endpoint of condensation as the
endpoint of the RG flow induced by a relevant perturbation
\cite{GNS:D0D2,KHM:rel}.  On the other hand, we can move
to a line of marginal stability where the vertex operator
becomes marginal and (if it is exactly marginal) 
study the deformation entirely in the CFT.

Consider the open strings between complexes given by $\cE^\bullet$ and
$\cF^\bullet$.
As in section \ref{ss:def} we will consider
an operator of ghost number one
$\phi\in\Hom^1(\cE^\bullet,\cF^\bullet)$. What happens to the
topological field theory if we use $\phi$ as a deformation?

Using exactly the same argument as in section \ref{ss:def} we find
that the resulting topological field theory has a single
D-brane consisting of the complex
\begin{equation}
\xymatrix@1@C=15mm{
\ar[r]&\cE^0\oplus\cF^0\ar[r]^{\left(\begin{smallmatrix}d_E&0\\
\phi&d_F \end{smallmatrix}\right)}&
\cE^1\oplus\cF^1\ar[r]^-{\left(\begin{smallmatrix}d_E&0\\
\phi&d_F \end{smallmatrix}\right)}&
\cE^2\oplus\cF^2\ar[r]&\ .
}  \label{eq:tachon}
\end{equation}
This is nothing other than $\Cone(\phi:\cE^\bullet[-1]\to\cF^\bullet)$.

At a general value of $(B+iJ)$ the operator $\phi$ may or may not 
represent a tachyon in the untwisted
theory. The mass of the spacetime field associated to
$\phi$ depends upon $B+iJ$; so the 
topological theory is independent of the mass.
If $\phi$ happens to be be
tachyonic then $\Cone(\phi:\cE^\bullet[-1]\to\cF^\bullet)$ represents
a ``bound state'' of $\cE^\bullet[-1]$ and $\cF^\bullet$. This is the
basic idea behind the ``$\Pi$-stability'' of \cite{DFR:stab,Doug:DC}.
Note that there are other formulations of stability as in
\cite{OPW:stab} for example.

Let us consider as an example the case of a \CY\ threefold $X$
and a particular D-brane/anti-D-brane pair. 
The D-brane
$\cF^\bullet$ is the simple, one-term complex
\begin{equation}
\xymatrix@1{
\ar[r]&0\ar[r]&\underline{\O_X}\ar[r]&0\ar[r]&\ ,
}  \label{eq:OX}
\end{equation}
where $\O_X$ is the structure sheaf of $X$.
This complex represents the trivial line bundle, or
basic D6-brane, over $X$. In
our notation the 
underline represents the location of the zeroth position in the
complex.

Let the anti-D-brane $\cE^\bullet$ be given by
\begin{equation}
\xymatrix@1{
\ar[r]&0\ar[r]&\O_X(-D)\ar[r]&\underline{0}\ar[r]&\ ,
}
\end{equation}
where $D$ is a particular 4-cycle $D\subset X$. That is, we have an
anti-D6-brane with a D4-brane charge given by $D$. (This is an
anti-brane rather than a brane because it appears in an odd position
in the complex.)

We now have a map $f:\O_X(-D)\to\O_X$ in
$\Hom^1(\cE^\bullet,\cF^\bullet)$ giving a cone:
\begin{equation}
\xymatrix@1{
\ar[r]&0\ar[r]&\O_X(-D)\ar[r]^-f&\underline{\O_X}\ar[r]&\ ,
}
\end{equation}
which is quasi-isomorphic to
\begin{equation}
\xymatrix@1{
\ar[r]&0\ar[r]&\underline{\O_D}\ar[r]&0\ar[r]&\ ,
}
\end{equation}
where $\O_D$ is the structure sheaf of $D$ extended by zero over $X$.

This is the topological field theory statement of a
D6-brane/anti-D6-brane pair forming a D4-brane state. Again let
us emphasize that the choice of grading above for this pair would
imply unbroken supersymmetry and hence marginal stability. Typically,
if $X$ is large, this will not be the case and $f$ will be a tachyon
which in turn makes the D4-brane a bound state.
At a given value of $(B+iJ)$ the {\it stable} object will
be described by a particular tachyon vev.

Finally, let us discuss the higher $\Ext$'s of section
\ref{ss:def}, which we chose to ignore at that point. 
For example, consider
$\Ext^2(\cE^n,\cE^{n-1})$. Writing the sheaves $\cE^n$ and $\cE^{n-1}$
as one-term complexes we may write
$\Ext^2(\cE^n,\cE^{n-1})=
\Hom^1(\underline{\cE}^n,\underline{\cE}^{n-1}[1])$. Thus,
even though $\Ext^2(\cE^n,\cE^{n-1})$ gives distinct 
deformations for a particular topological field theory, once we pass to
the big category $\CT(X)$, there are enough objects
(e.g. $\underline{\cE}^{n-1}[1]$) for us to replace all higher $\Ext$'s by
the deformations of the form we considered in (\ref{eq:tachon}). That
is, at least to first order, these higher $\Ext$'s do not deform the
theories outside the class given by $\CT(X)$.


\section{Zero-Brane Stability}   \label{s:D0}

In this section we will discuss 0-brane stability based on some
observations about monodromy. The basic idea is as follows. Let us
begin with a 0-brane which, we assume, is a stable object on large
\CY\ threefold. Now follow this object around a loop in the moduli
space of complexified K\"ahler form on $X$. If the resulting monodromy
results in something which is manifestly unstable then we must have
crossed a line of marginal stability for the 0-brane during our
travels, resulting in 0-brane decay.

This situation is similar to that in $N=2, d=4$
$SU(2)$ Yang-Mills theory \cite{SW:I}.
There the charge spectrum of stable BPS states
at weak coupling is not invariant under the monodromy
group of the theory.  If we follow the theory around
a loop which generates one of the offending
monodromy transformations, the BPS spectrum jumps.

In this case we can make a finer statement.  It may be
that states with D0-brane charge continue to
exist.  But there are many objects in $\DC(X)$ which
carry D0-brane charge.  Only some of these objects
are 0-branes in the sense of being points in a
large-volume Calabi-Yau.  We will ask about
the action of monodromy on these ``point'' objects.


\subsection{What is a 0-brane?}   \label{ss:D0}

Define $\underline{\cF}$ as a complex
containing all zeroes except $\cF$ at the zeroth position.
In the language of derived categories, a 0-brane on $X$ will be an
object in $\DC(X)$ which can be represented by $\underline{\O_p}$,
where $\O_p$ is the skyscraper sheaf of a point $p\in X$. Obviously $p$
is the location of the 0-brane in $X$.

Locally, in affine coordinates, we may use a Koszul resolution in
terms of free sheaves to 
represent the same object as:
\begin{equation}
\xymatrix@1@C=17mm{
0\ar[r]&
\O\ar[r]^-{\left(\begin{smallmatrix}z\\x\\y
\end{smallmatrix}\right)}&
\O^{\oplus3}\ar[r]^{\left(\begin{smallmatrix}y&0&-z\\
-x&z&0\\0&-y&x\end{smallmatrix}\right)}&
\O^{\oplus3}\ar[r]^-{\left(\begin{smallmatrix}x&y&z
\end{smallmatrix}\right)}&
\underline{\O}\ar[r]&0\ ,
}
\end{equation}
where $p$ is at $(x,y,z)=(0,0,0)$.

Now given any object $\cF^\bullet\in\DC(X)$ we may compute the
``D-brane charge'' in $H^{\textrm{even}}(X)$ in terms of a locally
free resolution as
\begin{equation}
  \ch(\cF^\bullet) = \sum_n(-1)^n\ch(\cF^n)\ .
\end{equation}
The D-brane charge of a 0-brane on a \CY\ threefold will be a 6-form
in De Rham cohomology which is Poincar\'e dual to a point.
The converse of this statement need not be true. There are
many objects in $\DC(X)$ which have the D-brane charge of a 0-brane but are
not quasi-isomorphic to a complex containing just $\O_p$.

Thus there is more to being a 0-brane than simply having 0-brane
charge.  Let us discuss a particularly clear illustration from a
{\em flop\/}, based on the work of Bridgeland \cite{Brig:flop} and also
noted in \cite{Doug:DC}.

Consider a \CY\ threefold $X$ and a flop $X'$ of $X$. It was shown
explicitly in \cite{Brig:flop} that there is an equivalence of
categories $\DC(X)\sim\DC(X')$. Assuming $X$ and $X'$ are
topologically inequivalent there is clearly no map equating the
0-branes of $X$ with the 0-branes of $X'$. Actually the
identification must be done as follows. Away from the exceptional
locus of the flop we may identify 0-branes of $X$ with 0-branes of
$X'$. There are then objects in $\DC(X)$ which correspond to
0-branes on the exceptional curve $C\subset X$ and there is a whole
bunch of different objects in $\DC(X)$ which correspond to 0-branes on
the other exceptional curve $C'\subset X'$.

That is, there is a plethora of objects in $\DC(X)$ which look like
they might be 0-branes. One way of distinguishing them
is via stability. The 0-branes corresponding to point on $C^\prime$ are
presumably unstable when $X$ is at large radius limit and {\em vice versa}.


\subsection{Monodromy}  \label{ss:mon}

Now consider the action of monodromy on objects in $\DC(X)$ as we
follow loops in the moduli space of the complexified K\"ahler form
$B+iJ$. The topological field theory of section \ref{s:DX} is
invariant under changes of the K\"ahler form and so the action of the
monodromy is manifestly trivial!

In order to see monodromy we need to restore the dependence
of the grading on $(B+iJ)$. Now if we identify a
given physical D-brane as an object in $\DC(X)$, the monodromy will act
nontrivially to produce a different object in $\DC(X)$. At this point
we do not understand how to see the action of this monodromy directly.  
We will discuss some hints for how it arises in section \ref{ss:phys}.

Here we will instead assume a conjecture by Kontsevich, Morrison and Horja
\cite{Kont:mon,Mor:geom2,Horj:DX,me:navi}. At least for \CY\
threefolds embedded as complete intersections in toric varieties there
is a distinguished divisor in the moduli space known as the ``primary
component of the discriminant'' \cite{Horj:DX,me:navi}.\footnote{Also
known sometimes as the ``principal'' component.} It seems reasonable
to assume that this component of the discriminant can be defined for
general \CY\ threefolds.

According to the conjecture by Kontsevich et al, a loop around the
primary component of the discriminant transforms an element of
$\DC(X)$ as
\begin{equation}
T(\cF^\bullet) = \Cone\Bigl(\hom(\underline{\O_X},\cF^\bullet)
\otimes\underline{\O_X}\to\cF^\bullet\Bigr)\ ,
\label{eq:principalm}
\end{equation}
where the operator ``$\otimes$'' is defined in the obvious
way in the derived category and is explained further, along with many
other interesting facts about this Fourier--Mukai transform, in
\cite{ST:braid}. A generalization of this Fourier--Mukai transform is
discussed in \cite{Horj:EZ}.

What happens to a 0-brane upon monodromy around the primary
component? First note that if $\cF^\bullet$ is of the form
$\underline{\cF}$ (i.e., concentrated at position zero) then the
cohomology of the complex
$\hom(\underline{\O_X},\cF^\bullet)$ is given by the sheaf cohomology
of $\cF$. The skyscraper sheaf has trivial cohomology given by
$H^0=\C$, all other cohomology vanishes. It follows that
$\hom(\underline{\O_X},\underline{\O_p})\otimes\underline{\O_X}$ is
given simply by $\underline{\O_X}$. Therefore
\begin{equation}
T(\underline{\O_p}) = \Bigl(\O_X\to\underline{\O_p}\Bigr)\ ,
        \label{eq:D01}
\end{equation}
which is quasi-isomorphic to $\underline{\cI_p}[1]$ --- the ideal sheaf of a
point $p$ shifted left by one.


\subsection{Stability}  \label{ss:stab}

Stability of the 0-brane at small volume now depends
on whether $\underline{\cI_p}[1]$ can become a stable object
in the large radius limit.  By this we mean that
either the anti-D6-brane $\O_X$ and the D0-brane $\O_p$
are mutually supersymmetric, or they can form
a supersymmetric bound state (as with the D0-D2 system).
A signature of the latter would be an attractive force
between the objects at short distance.

We can ask about stability at arbitrarily large volume;
the force is then identical to that between the
D0 and anti-D6 branes in flat space.

The static force is computed for the D0-D6 system in
\cite{Lif:loop,Shein:06}, and is repulsive.  The answer is identical
for the D0/anti-D6 system.  The potential energy is equal to the
one-loop vacuum energy for open strings stretched between the two
branes.  Reversing the charge of the D6-brane in this calculation
simply changes the parity $(-1)^F$ of open string states preserved by
the GSO projection.  But due to fermion zero modes in the D0/anti-D6
system, $\tr (-1)^F q^{L_0}$ vanishes in both the Neveu-Schwarz and
Ramond sectors \cite{Lif:loop}; so the one-loop amplitude is
independent of the GSO projection.  Therefore the D0- and
anti-D6-branes also repel each other.  Another argument is that the
two D-branes break spacetime supersymmetry at large volume
\cite{Wit:D0D6} (see also the earlier paper
\cite{MPT:Dsol}).\footnote{In \cite{Wit:D0D6} a stable state was
found, but at large NS-NS B-field.  We may avoid this issue by
beginning and ending at $\Rea (B + i J) = 0$.  It would be interesting
to see if and how the result in \cite{Wit:D0D6}\ applies to stability
issues on the Calabi-Yau threefold.}

Using the same logic as in \cite{SW:I},
we see that if one follows a loop around the primary component
of the discriminant locus then we must cross a line of marginal
stability for any 0-brane and hence the 0-branes will be unstable for
some subset of the loop.

The location of the primary component of the discriminant was
discussed in \cite{me:navi}. If one begins at large radius then a loop
around the primary component will necessarily leave the
``geometric phases'' region of the moduli space. For example, in the
simple case of the quintic threefold, the primary component lies in
the phase boundary between the large-radius \CY\ phase and the \LG\
phase. In more complicated threefold examples, the primary component
will appear 
in the wall of a phase boundary whenever the effective dimension of
the target space decreases below three. For example, one will {\em
not\/} see the primary component when blowing down a divisor to
produce an orbifold singularity but one can see the primary component
if one passes to a ``hybrid'' phase of a \LG\ theory fibred over
$\P^2$ say. One also passes to a non-geometric phase when performing the
``exoflop'' of \cite{AGM:II}.

This shows that the paths associated with destabilizing the 0-brane pass
outside the geometric phases. 
To determine the exact way in which the 0-brane becomes unstable
requires further analysis which we do not do here.

There may well be objects which are stable in a
non-geometric phase, such as a \LG\ phase, and have the right charge to
be a 0-brane but are not a 0-brane in the strict sense of section
\ref{ss:D0}. This could explain the ``zero brane''
found at the Gepner point in 
\cite{DR:Dell}.  Note that even if the
three massless fields found in \cite{DR:Dell}\
are moduli, the moduli space may be different
from the large-volume threefold.

Another explanation of the D0-brane in \cite{DR:Dell}\
is that the D0-brane really remains stable near the
Gepner point.  It would be interesting to 
understand this further.

We note in passing that the conjectured Matrix theory
description of Calabi-Yau compactifications of M-theory
is as a collection of $N\to\infty$ BPS objects with
0-brane charge
at the primary component of the discriminant \cite{KLS:CY}.
The instability of the ``point-like'' 0-brane
at this component of the moduli space
is highly relevant to the dynamics of this large-N theory.


\subsection{Another time around\protect\footnote{We are very grateful to
S.~Katz for conversations regarding this section.}}
   \label{ss:again}

It is interesting to ask what happens if we take our 0-brane around
the primary component a {\em second\/} time. That is, what
does (\ref{eq:principalm}) give for
$T(\underline{\cI_p}[1])$?

From the exact sequence $0\to\cI_p\to\O_X\to\O_p\to0$ it is easy to compute
that $H^3(X,\cI_p)=\C$ with all other cohomology groups
vanishing. This implies that
\begin{equation}
\hom(\underline{\O_X},\underline{\cI_p}[1])\otimes
\underline{\O_X}=\underline{\O_X}[-2]\ .
\end{equation}
Now let $I^\bullet$ be an injective resolution of $\cI_p$. It follows
that
$T(\underline{\cI_p}[1])=\Cone(\underline{\O_X}[-2]\to\underline{\cI_p}[1])$
is given by
\begin{equation}
\xymatrix@R=0pt{
I^0\ar[r]^{i_0}&\underline{I^1}\ar[r]^{i_1}&I^2\ar[r]^{i_2}&I^3\ar[r]^{i_3}&\\
&&\oplus&&\\
&&\O_X\ar[uur]_f&&
}  \label{eq:D0twice}
\end{equation}
where $f$ is a map which generates the sheaf cohomology group
$H^3(X,\cI_p)$.

The complex (\ref{eq:D0twice}) represents the weird and
wonderful object in $\DC(X)$ obtained by looping a 0-brane twice around the
primary component of the discriminant.
This complex represents one of the interesting objects in $\DC(X)$
which cannot be reduced to its cohomology. To see this note first that
the cohomology of (\ref{eq:D0twice}) is the same as the cohomology of
\begin{equation}
\xymatrix{
0\ar[r]^0&\cI_p\ar[r]^0&\underline{0}\ar[r]^0&\O_X\ar[r]^0&\ .
}  \label{eq:mun}
\end{equation}
One might therefore be tempted to say that (\ref{eq:D0twice})
represents an anti-D6-brane/D0-brane pair at position $-1$ (giving
$\underline{\cI_p}[1]$) together with an an anti-D6-brane at position
$1$ (giving $\underline{\O_X}[-1]$). It turns out however that 
(\ref{eq:D0twice}) is {\em not\/} quasi-isomorphic to
(\ref{eq:mun}). One may check this by computing $\Hom^P(\O_X,-)$ for
each complex for example. Thus the full glory of the object
(\ref{eq:D0twice}) cannot be reduced to a statement of conventional
branes.

Since our na\"\i ve picture of branes as subspaces, and hence coherent
sheaves, is tied to large radius pictures, one might be tempted to
speculate that exotic objects such as (\ref{eq:D0twice}) will not be
stable at large radius limits. Certainly we expect this to be true in
this case since our 0-brane already decayed after one trip around the
primary component of the discriminant.

It would be interesting to prove the instability at large radius of
these objects which defy a simple D-brane interpretation.


\section{Further Discussion}    \label{s:disc}

\subsection{Relation of topological to physical branes}  \label{ss:phys}

We have emphasized that at a general point
in the K\"ahler moduli space, a general object
in $\DC(X)$ will {\em not} correspond to
a physical D-brane, and the topological
field theory will not be a twist of a CFT
at that point.  This is in line with the
ideology in \cite{Doug:DC}\ that
when the D-branes are taken to fill our four-dimensional
spacetime and so realize $N=1$ compactifications, the
objects in $\DC(X)$ correspond to solutions to the
F-term equations, while the stability conditions
are essentially the D-term equations.

At the line of marginal stability, the tachyonic
deformations of a collection of branes become
marginal.  If they are exactly marginal we
may pick any deformation we like along the
marginal directions and still have a CFT.
We may twist this CFT to get a topological field theory with
arbitrary deformations, as we have described.

As we move away from this line, the topological field theory
is invariant.  The flow of grading means the
associated deformations will be tachyonic
(dependig on which direction we move
off the line of marginal stability).
At a given value of $B+iJ$, only a subset 
of measure zero of the
tachyon vevs will correspond
to CFTs.  This subset corresponds to the tachyon
at the extrema of its effective potential,
or to some stable solitonic configuration of the
tachyon.  The remaining topological field theories will describe
``off-shell'' configurations of string theory.

The change of grading indicates that the
tachyon potential (and thus its extrema)
will vary throughout the K\"ahler moduli space.
Therefore the stable conformal field theories will
change, giving some notion of a nontrivial ``bundle''
of CFT data over K\"ahler moduli space.
This change should lead to monodromy action
on $\DC(X)$, at least on states in $\DC(X)$ which exist as
physical objects at our starting point in K\"ahler
moduli space.


\subsection{Linear sigma models for complexes}

There are now constructions via linear sigma
models (following \cite{W:phase}) of a class of
D-branes described by monads 
\cite{lsm:monad,Hori:lsm,GJ:lsmseq}\ and by 
multi-step resolutions \cite{lsm:multistep,GJ:lsmseq} in toric
varieties.\footnote{Previous
work on open string linear sigma models includes
\cite{HIV:D-mir, GJS:lsm}.} Some of the structures
in \cite{Doug:DC}\ and in the present work
appear quite naturally and intuitively in the linear sigma
model.  The Chan-Paton
factors are zero modes of boundary fields.
The chain maps are mass terms pairing up the
fields of neighboring complexes, and their
appearance via boundary contributions
to the worldsheet supercharge is clear.

Since one may pass between phases easily in the linear
sigma model,
it may be useful to study 0-brane
decay in this framework.  Hopefully using this
framework, or that advocated in \cite{DD:stringy},
one may be able to understand what the 0-brane decays
to, and what if any object at small radius might
have 0-brane charge.


\subsection{$\DC(X)$ as a birational invariant} \label{ss:birat}

Suppose  $X$ and $X'$ are birationally equivalent
projective \CY\ threefolds. For simplicity we assume $X$ and $X'$ are
smooth although this constraint can almost certainly be relaxed to
some extent. It was shown in 
\cite{BO:DCeq,Brig:flop} that $\DC(X)$ is equivalent to $\DC(X')$.
We may use our topological field theory to motivate a
generalization of this kind of statement. 

Let $X$ be a \CY\ variety of dimension $d$. The phase picture of
\cite{AGM:II} shows how, for the case $d=3$, a flop to $X'$ may be seen as a
change in $B+iJ$. The basic idea is that if you shrink a $\P^1$ down
to zero size by changing the K\"ahler form, continuing this process
through the wall of K\"ahler cone will result in a flop on this curve.
Thus $X$ and $X'$ are related simply by a change in
the K\"ahler structure. 

Now the notion of D-branes should be intrinsic to the worldsheet. That
is, we should not need a manifest geometric realization of the target
space geometry in order to understand D-branes. One should therefore view
$\CT(X)$ as being constructed from an underlying worldsheet
picture rather than being explicitly computed from $X$. 
Since $\DC(X)\sim\CT(X)$ is invariant
under a change in the worldsheet data that manifests itself as
a change in $B+iJ$, it follows immediately that $\DC(X)\sim\DC(X')$!

Note also that this can be generalized to any $d$. If
a birational transformation between \CY's can be induced by a change
in the K\"ahler 
structure then the derived category will be invariant. It would be
interesting to know if all birational transformations between \CY's
can be induced this way for $d>3$. 


\section*{Acknowledgments}

It is a pleasure to thank M.~Douglas, S.~Hellerman, S.~Katz, J.~McGreevy,
D.~Morrison, R.~Plesser, L.~Saper,
E.~Sharpe and M.~Vybornov for useful conversations. 
P.S.A.~is supported in part by NSF grant DMS-0074072 and a research
fellowship from the Alfred P.~Sloan Foundation. 
A.L.~is supported by the Department of Energy under
contract DE-AC03-76SF00098.  A.L.~would like
to thank the Center for Geometry and Theoretical Physics
at Duke University for their hospitality during the
course of this project. 


\end{document}